\documentclass[revtex4,pre, preprint, epsfig, showpacs, superscriptaddress]{revtex4}

\usepackage[dvips]{graphicx}
\usepackage{latexsym}
\usepackage{natbib}

\begin{document}

\bibliographystyle{apsrev}

\title{Betweenness Centrality of Fractal and Non-Fractal Scale-Free Model Networks and
  Tests on Real Networks}

\author{Maksim Kitsak}
\affiliation{Center for Polymer Studies, Boston University, Boston,
  Massachusetts 02215, USA}

\author{Shlomo Havlin}
\affiliation{Center for Polymer Studies, Boston University, Boston,
  Massachusetts 02215, USA}
\affiliation{Minerva Center and Department of Physics, Bar-Ilan
  University, Ramat Gan, Israel}

\author{Gerald Paul}
\affiliation{Center for Polymer Studies, Boston University, Boston,
  Massachusetts 02215, USA}

\author {Massimo Riccaboni}
\affiliation{Faculty of Economics, University of Florence, Via
delle Pandette 9, Florence, 50127, Italy}

\author{Fabio Pammolli}
\affiliation{Faculty of Economics, University of Florence, Via delle
Pandette 9, Florence, 50127, Italy}

\affiliation{IMT Institute for Advanced Studies, Via S. Micheletto
3, Lucca, 55100, Italy}

\author{H. Eugene Stanley}
\affiliation{Center for Polymer Studies, Boston University, Boston,
  Massachusetts 02215, USA}

\date{\today (printed) -- khpps.tex -- last revised:14 Feb. 2007}

\begin{abstract}

We study the betweenness centrality of fractal and non-fractal
scale-free network models as well as real networks.  We show that
the correlation between degree and betweenness centrality $C$ of
nodes is much weaker in fractal network models compared to
non-fractal models. We also show that nodes of both fractal and
non-fractal scale-free networks have power law betweenness
centrality distribution $P(C)\sim C^{-\delta}$. We find that for
non-fractal scale-free networks $\delta = 2$, and for fractal
scale-free networks $\delta = 2-1/d_{B}$, where $d_{B}$ is the
dimension of the fractal network. We support these results by
explicit calculations on four real networks: pharmaceutical firms
($N=6776$), yeast ($N=1458$), WWW ($N=2526$), and a sample of
Internet network at AS level ($N=20566$), where $N$ is the number
of nodes in the largest connected component of a network. We also
study the crossover phenomenon from fractal to non-fractal
networks upon adding random edges to a fractal network. We show
that the crossover length $\ell^{*}$, separating fractal and
non-fractal regimes, scales with dimension $d_{B}$ of the network
as $p^{-1/d_{B}}$, where $p$ is the density of random edges added
to the network. We find that the correlation between degree and
betweenness centrality increases with $p$.

\end{abstract}

\pacs{89.75.Hc}

\keywords{Keywords}

\maketitle

\section{Introduction}

Studies of complex networks have recently attracted much attention
in diverse areas of science \cite{Barabasi_rmp_review, evol_of_nw,
evol_and_struct, compl_nw}. Many real-world complex systems can be
usefully described in the language of networks or graphs, as sets
of nodes connected by edges \cite{ER, ER_random graphs}. Although
different in nature many networks are found to possess common
properties. Many networks are known to have a ``small-world''
property \cite{collect_dynam, sw_problem, bollobas, diam_of_www}:
despite their large size, the shortest path between any two nodes
is very small. In addition, many real networks are scale-free (SF)
\cite{Barabasi_rmp_review, evol_of_nw,evol_and_struct, compl_nw,
faloutsos_etall, scaling_in_nw}, well approximated by a power-law
tail in degree distribution, $ P(k)\sim k^{-\lambda} $, where $k$
is the number of links per node.

Many networks, such as the WWW and biological networks have
self-similar properties and are fractals \cite{repulsion,self-sim,
skeleton_and_fractal,crit_and_supercrit_skelet, dissasort}. The
box-counting algorithm \cite{box_count, self-sim} allows to
calculate their fractal dimensions $d_{B}$ from the box-counting
relation
\begin{equation}
N_{B}\sim \ell_{B}^{-d_{B}}, \label{dim}
\end{equation}
where $N_{B}$ is the minimum number of boxes of size $\ell_{B}$
needed to cover the entire network (Appendix B). Structural
analysis of fractal networks shows that the emergence of SF
fractal networks is mainly due to disassortativity or  repulsion
between hubs \cite{repulsion}. That is, nodes of large degree
(hubs) tend to connect to nodes of small degree, giving life to
the paradigm ``the rich get richer but at the expense of the
poor.'' To incorporate this feature, a growth model of SF fractal
networks that combines a renormalization growth approach with
repulsion between hubs has been introduced \cite{repulsion}. It
has also been noted \cite{repulsion} that the traditional measure
of assortativity of networks, the Pearson coefficient $r$
\cite{assortativity} does not distinguish between fractal and
non-fractal network since it is not invariant under
renormalization.

Here, we study properties of fractal and non-fractal networks,
including both models and real networks. We focus on one important
characteristic of networks, the {\it betweenness centrality\/}
(C), \cite{centr_definition, centr_book, centr_handbook, newman},
defined as,
\begin{equation}
C(i)\equiv \sum_{j,k}\frac{ \sigma_{j,k}(i) }{\sigma_{j,k}},
\label{centrality}
\end{equation}
where $\sigma_{j,k}(i)$ is the number of shortest paths between
nodes $j$ and $k$ that pass node $i$ and $\sigma_{j,k}$ is the total
number of shortest paths between nodes $j$ and $k$.

The betweenness centrality of a node is proportional to the number
of shortest paths that go through it. Since transport is more
efficient along shortest paths, nodes of high betweenness centrality
$C$ are important for transport. If they are blocked, transport
becomes less efficient. On the other hand, if the capacitance of
high $C$ nodes is improved, transport becomes significantly better
\cite{superhighways}.

Here we show that fractal networks possess much lower correlation
between betweenness centrality and degree of a node compared to
non-fractal networks. We find that in fractal networks even small
degree nodes can have very large betweenness centrality while in
non-fractal networks large betweenness centrality is mainly
attributed to large degree nodes. We also show that the
betweenness centrality distribution in SF fractal networks obeys a
power law. We study the effect of adding random edges to fractal
networks. We find that adding a small number of random edges to
fractal networks significantly decreases the betweenness
centrality of small degree nodes. However, adding random edges to
non-fractal networks has a significantly smaller effect on the
betweenness centrality.

We also analyze the transition from fractal to non-fractal
networks by adding random edges and show both analytically and
numerically that there exists a crossover length $\ell^{*}$ such
that for length scales $\ell < \ell^{*}$ the topology of the
network is fractal while for $\ell
>\ell^{*}$ it is non-fractal. The crossover length scales as
$\ell^{*}\sim p^{-1/d_{B}}$, where $p$ is the number of random
edges per node.  We analyze seven SF model networks and four real
networks.

The four real networks we analyze are the network of
pharmaceutical firms \cite{Pharm}, an Internet network at the AS
level obtained from the DIMES project \cite{dimes, Shai}, PIN
network of yeast \cite{Barabasi, yeast} and WWW network of
University of Western Sydney \cite{uws}. Pharmaceutical network is
the network of nodes representing firms in the worldwide
pharmaceutical industry and the links are collaborative agreements
among them. The Internet network represents a sample of the
internet structure at the Autonomous Systems(AS) level. The
Protein Interaction Network (PIN) of yeast represents proteins as
nodes and interactions between them as links between nodes. The
WWW network of University of Western Sydney represents web pages
(nodes) targeted by links from the uws.edu.au domain. Basic
properties of the considered networks are summarized in Table. I.

The manuscript is organized as follows: In section II, we study
correlation between the betweenness centrality and degree of
nodes, and we compare fractal and non-fractal networks. We analyze
the betweenness centrality variance $\sigma_{C}(k)$ of nodes of
the same degree $k$ and introduce a correlation coefficient $R$
that describes the strength of betweenness centrality degree
correlation. We also analyze the betweenness centrality
distribution $P(C)$ of several model and real networks. In section
III we study the transition from fractal to non-fractal networks
with randomly added edges. Appendix A provides a short summary of
the fractal growth model introduced in \cite{repulsion}. Appendix
B discusses the box covering method and its approximations.

\section{Betweenness centrality of fractal and non-fractal networks}

\begin{figure}[!ht]
\includegraphics[width=10.0 cm,height=9.0cm,angle=0]{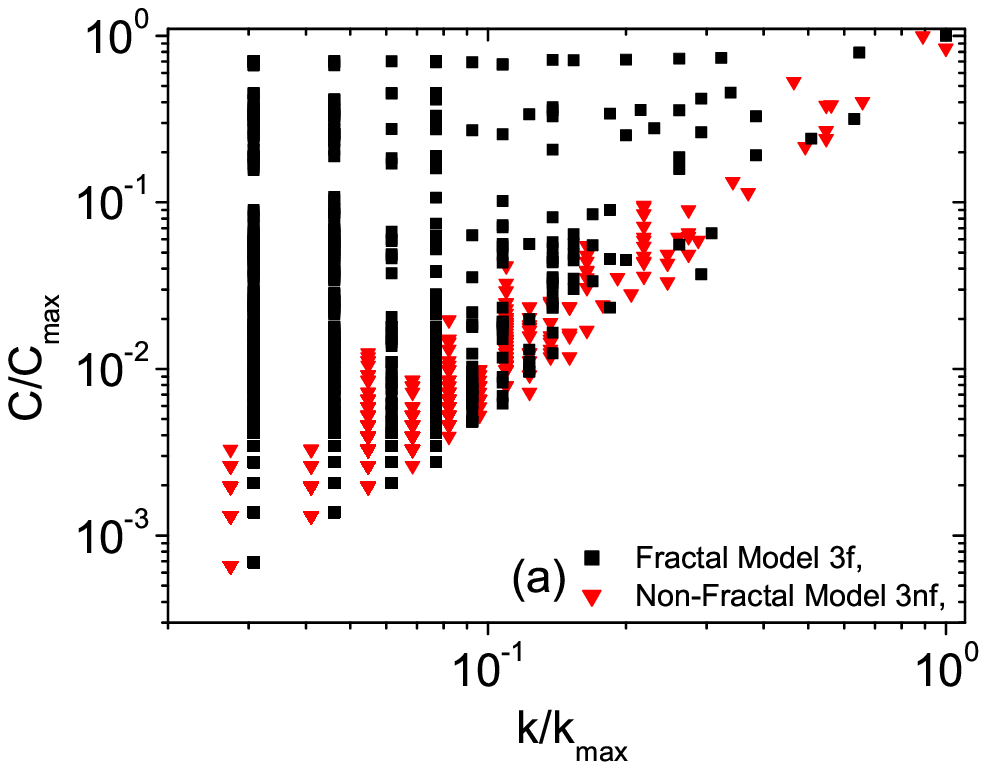}
\includegraphics[width=10.0 cm,height=9.0cm,angle=0]{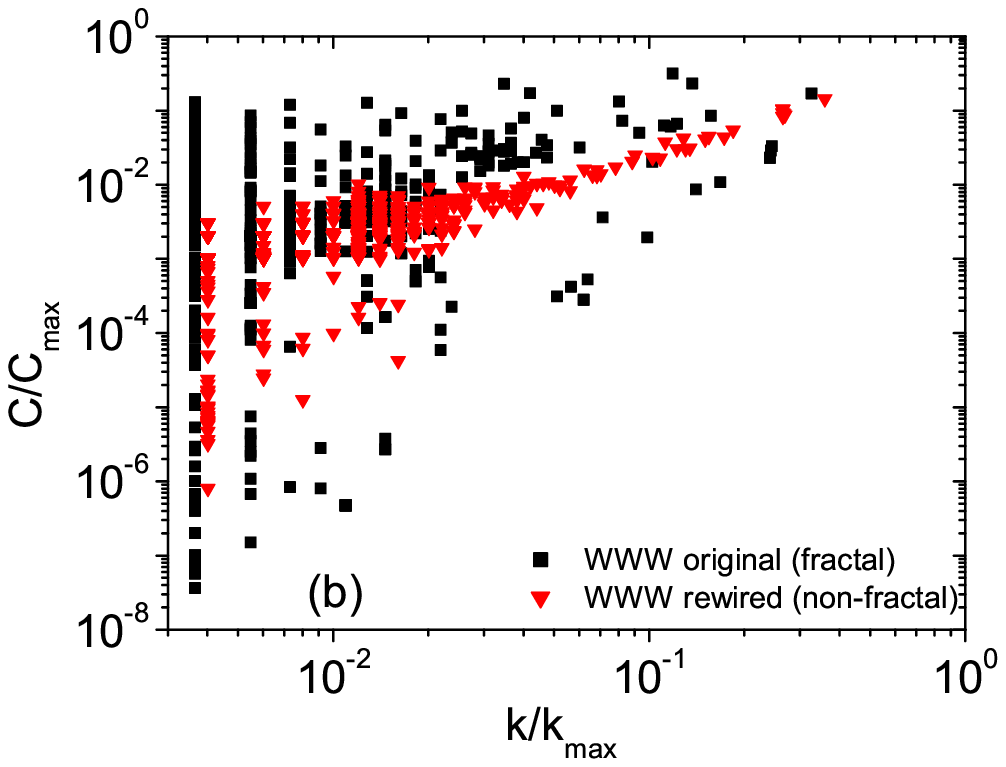}
\caption { (a) Betweenness centrality versus degree correlation
profiles of fractal and non-fractal network models. Note the
broader range of betweenness centrality values of small degree
nodes of fractal network compared to that of the non-fractal
network. (b) Betweenness centrality versus degree correlation
profiles of Uni Western Sydney WWW (fractal) network and its
random counterpart. The randomly rewired network is non-fractal.
Betweenness centrality and degree are correlated much stronger in
nodes of the random rewired network.} \label{fig_centr_vs_degree}
\end{figure}

It is generally accepted \cite{Attack_vulnerability} that in many
networks nodes having a larger degree also have a larger betweenness
centrality. Indeed, the larger the degree of a node, the larger the
chance that many of the shortest paths will pass through this node;
the chance of many shortest paths passing a low degree node is
presumably small. Here we show that this is not the case for fractal
SF networks. As seen in Fig.~\ref{fig_centr_vs_degree}(a) small
degree nodes in fractal SF networks have a broad range of
betweenness centrality values. The betweenness centrality of many
small degree nodes can be comparable to that of the largest hubs of
the network. For non-fractal networks, on the other hand, degree and
betweenness centrality of nodes are strongly correlated.

To demonstrate the difference in the relation between degree and
betweenness centrality in real networks we compare original
networks with their random (uncorrelated) counterparts. We
construct the random counterpart network by rewiring the edges of
the original network, yet preserving the degrees of the nodes and
enforcing its connectivity. As a result we obtain a random network
with the same degree distribution which is \emph{always}
non-fractal regardless of the original network. As seen in
Fig.~\ref{fig_centr_vs_degree}(b),the betweenness
centrality-degree correlation of a random network obtained by
rewiring edges of the WWW network is much stronger compared to
that of the original network. Ranges of betweenness centrality
values for a given degree decrease significantly as we randomly
rewire edges of a fractal SF network.

The quantitative description of the betweenness centrality -
degree correlation can be given by the analysis of the betweenness
centrality variance $\sigma_{C}(k)$ attributed to nodes of
specific degree $k$. Larger values of the variance $\sigma_{C}(k)$
mean weaker correlations between degree and betweenness centrality
of a node since nodes of the same degree have larger variations in
betweenness centrality values. As seen in Fig.~\ref{variance
profiles}, in a region of small degree, betweenness centrality
variance $\sigma_{C}(k)$ of fractal networks is significantly
bigger than that of their respective randomly rewired counterparts
which are not fractals. At the same time betweenness centrality
variance of non-fractal networks is comparable or even smaller
than that of the corresponding randomly rewired networks. Thus,
the betweenness centrality of nodes of fractal networks is
significantly less correlated with degree than in non-fractal
networks.

This can be understood as a result of the repulsion between hubs
found in fractals \cite{repulsion}: large degree nodes prefer to
connect to nodes of small degree and not to each other. Therefore,
the shortest path between two nodes must necessarily pass small
degree nodes which are found at all scales of a network. Thus, in
fractal networks small degree nodes have a broad range of values
of betweenness centrality while in non-fractal networks nodes of
small degree generally have small betweenness centrality.
Betweenness centralities of small degree nodes in fractal networks
significantly decrease after random rewiring since the rewired
network is no longer fractal. On the other hand, centralities of
nodes in non-fractal networks either do not change or increase
after rewiring of edges.

As seen in Fig.\ref{fig_centr_vs_degree}(b), the main difference
in the betweenness centrality - degree correlation between fractal
and non-fractal SF networks reveals itself in the dispersion of
betweenness centrality values attributed to nodes of given degree,
rather than in the average betweenness centrality
values.\footnote{Due to the fact that the average betweenness for
a given degree doesn't change much, the Pearson coefficient, as a
traditional measure of correlation, is not suitable to
characterize the differences in betweenness centrality - degree
correlation of fractal an non-fractal networks. This is since the
Pearson coefficient is dominated by the average values of the
betweenness centrality for a given degree.} So, in order to
characterize and quantify the overall betweenness centrality -
degree correlation we propose a correlation dispersion
coefficient:
\begin{equation}
R = \frac {\sum_{k} \sigma_{C}(k)*p(k)} {\sum_{k}
\sigma^{*}_{C}(k)*p(k)}, \label{coeff}
\end{equation}
where $\sigma_{C}(k)$ and $\sigma^{*}_{C}(k)$ are the betweenness
centrality variances  of the original and randomly rewired
networks respectively and $p(k)$ is the degree distribution of
both networks. The dispersion coefficient $R$ is the ratio between
the mean variance $<\sigma_{C}(k)>$ of the original network and
$<\sigma^{*}_{C}(k)>$, the mean variance of the randomly rewired
network. We note that fractal SF networks have bigger values of
the betweenness centrality variance than their randomly rewired
counterparts and therefore, have correlation dispersion
coefficient $R>1$. On the other hand $\sigma_{C}(k)$ of the
non-fractal SF networks is close or smaller than that of their
random counterparts $\sigma^{*}_{C}(k)$ which result in values of
the correlation dispersion coefficient $R\approx1$ or $R<1$. The
calculated values of the correlation dispersion coefficient $R$
for the networks we considered in the paper are summarized in
Table. I.

\begin{figure}[!ht]
\includegraphics[width=10.0 cm,height=9.0cm,angle=0]{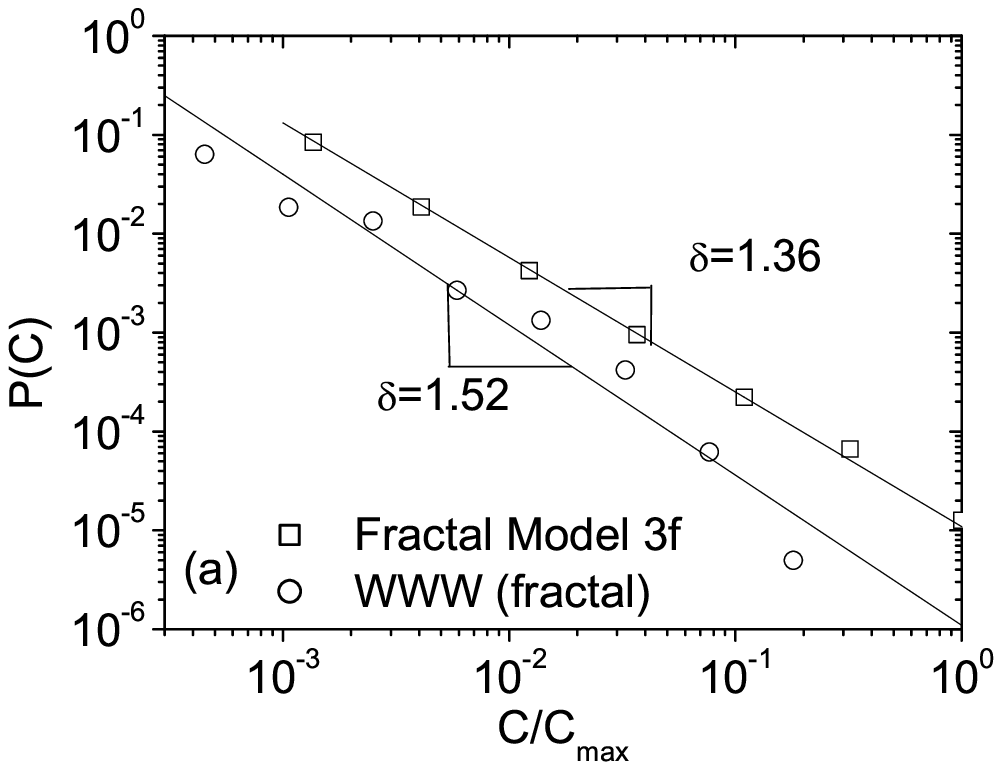}
\includegraphics[width=10.0 cm,height=9.0cm,angle=0]{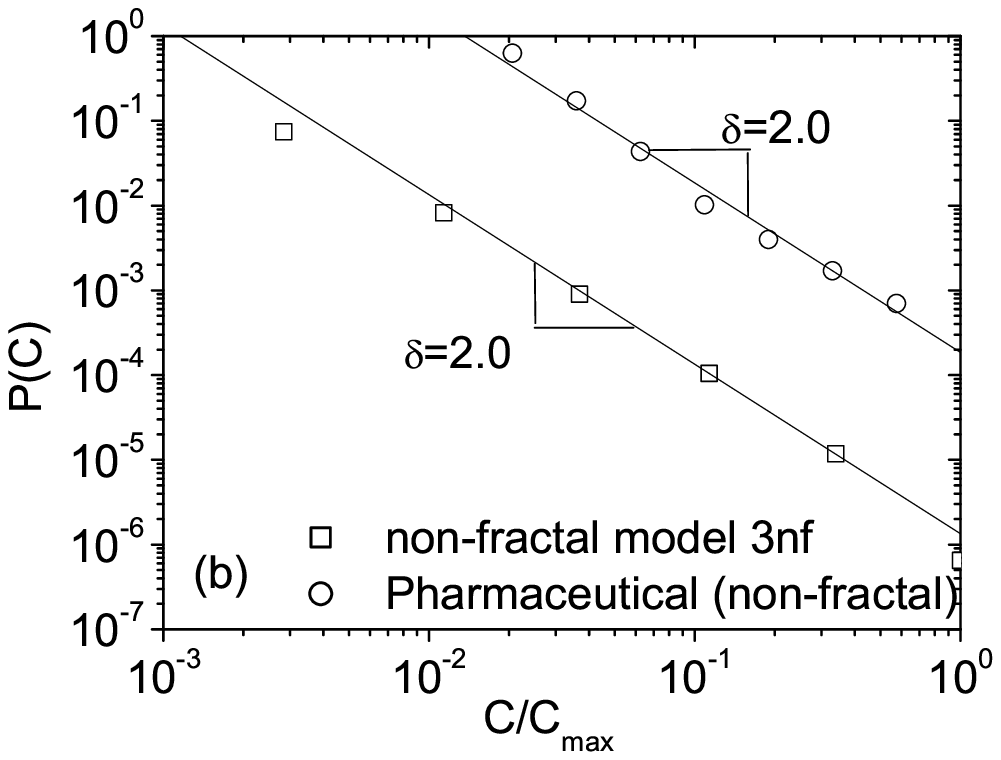}
\caption{Betweenness centrality distributions of (a) Fractal model
3f and WWW network (fractal) and (b) non-fractal model 3nf and
pharmaceutical network (non-fractal). The data have been binned
logarithmically. Both fractal and non-fractal networks exhibit a
power-law range of betweenness centrality distribution consistent
with $P(C) \sim C^{-\delta}$. The measured betweenness centrality
distributions (data points) are in good agreement with
analytically obtained formula $\delta = 2 - 1/d_{B}$ represented
by the straight lines. In non-fractal networks we expect $\delta
\rightarrow 2$ since $d_{B}\rightarrow \infty$.}
\label{fig_centr_distr}
\end{figure}

The probability density function (pdf) of betweenness centrality
has been studied for both Erd\"{o}s  R\'{e}nyi \cite{ER,ER_random
graphs} and SF \cite{scaling_in_nw} networks. It was found that
for SF networks the betweenness centrality distribution satisfies
a power law
\begin{equation}
P(C)\sim C^{-\delta}, \label{centr_distr}
\end{equation}
with typical values of $\delta$ between 1 and 2 \cite{centr_distr_1,
centr_distr_2, Braunstein_review}. Our studies of the betweenness
centrality distribution support these earlier results
(Fig.~\ref{fig_centr_distr}). We find that $\delta$ increases with
dimension $d_{B}$ of analyzed fractal networks. In the case of
non-fractal networks, where $d_{B}= \infty$, estimated values of
$\delta$ seem to be close to $2$.

An analytic expression for $\delta$ can be derived for SF fractal
tree networks by using arguments similar to those used in
\cite{Braunstein_review} to find $\delta$ for the minimum spanning
tree (MST). Consider a fractal tree network of dimension $d_{B}$.
A small region of the network consisting of $n$ nodes will have a
typical diameter $\ell(n)\sim n^{1/d_{B}}$ \cite{Bunde_Havlin}.
Nodes in this region will be connected to the rest of the network
via $\ell(n)$ nodes. Thus, the betweenness centrality of those
nodes is at least $n$. Since the number of regions of size $n$ is
$N/n$, the total number of nodes with betweenness centrality $C>n$
in the network is

\begin{equation}
\phi(C>n) \sim\ell(n){N\over n}\sim n^{1/d_{B} - 1}.
\end{equation}
Thus, the number of links with betweenness centrality $n$ is
\begin{equation}
P(C) = \Delta \phi \sim \phi(C+1) - \phi(C) \sim C^{1/d_{B} - 2}.
\end{equation}
Using Eq.~(\ref{centr_distr}) we immediately obtain
\begin{equation}
\delta = 2 - {1\over d_B} \label{centr_delta}.
\end{equation}
Thus, Eq.~(\ref{centr_delta}) shows that $\delta$ increases with
$d_{B}$ in agreement with Fig.~\ref{fig_centr_distr}. For
non-fractal networks $d_{B} \rightarrow \infty$ and $\delta = 2$.
So non-fractal networks consist of relatively small number of
central nodes and a large number of leaves connected to them. On
the other hand in fractal networks, especially in those of small
dimensionality, due to the repulsion between hubs, betweenness
centrality is distributed among all nodes of a network. Analysis
of the box covering method as a fractal test for some fractal and
non-fractal networks studied here is shown in
Fig.~\ref{fig_nw_dims}.

\section{Crossover scaling in fractal networks}

\begin{figure}[!ht]
\includegraphics[width=7.0 cm,height=5.0cm,angle=0]{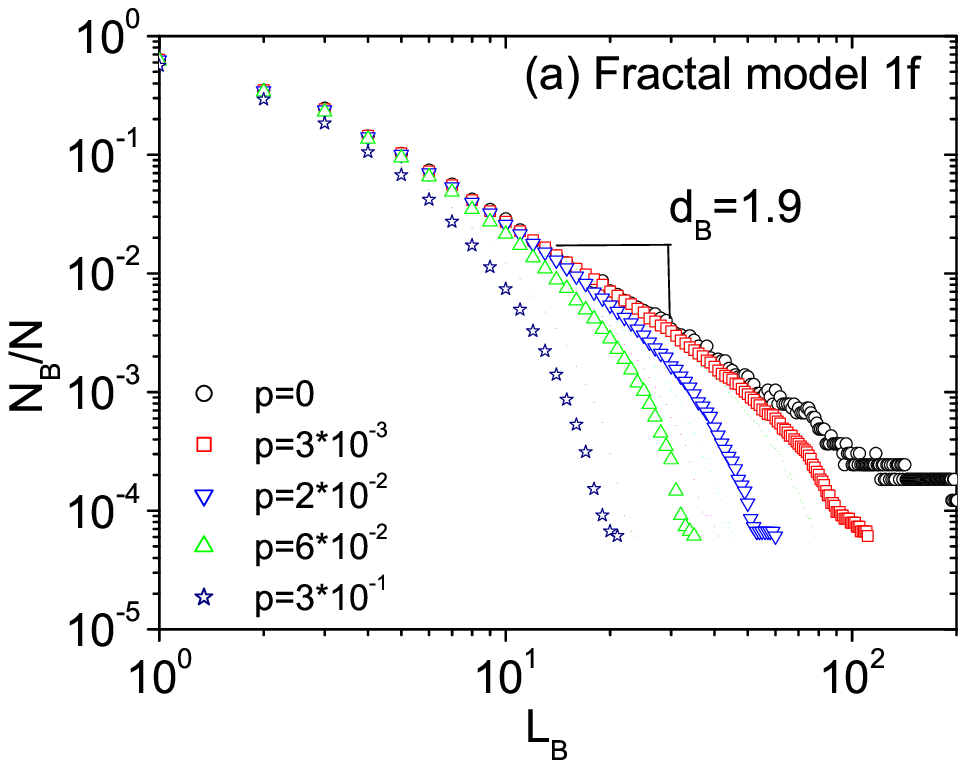}
\includegraphics[width=7.0 cm,height=5.0cm,angle=0]{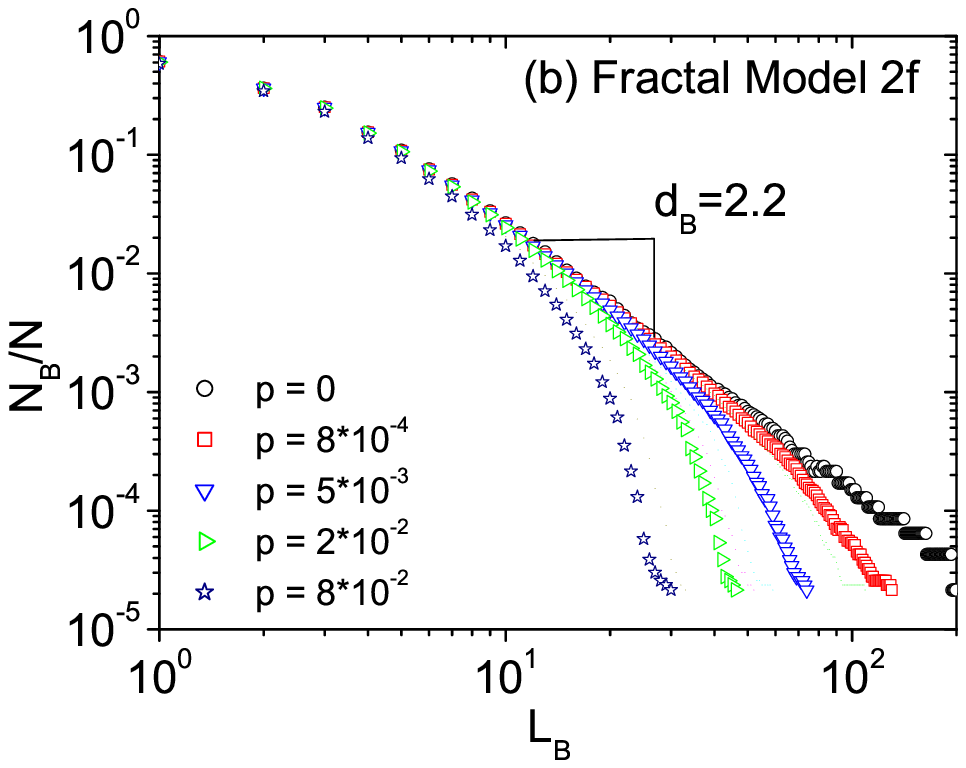}
\includegraphics[width=7.0 cm,height=5.0cm,angle=0]{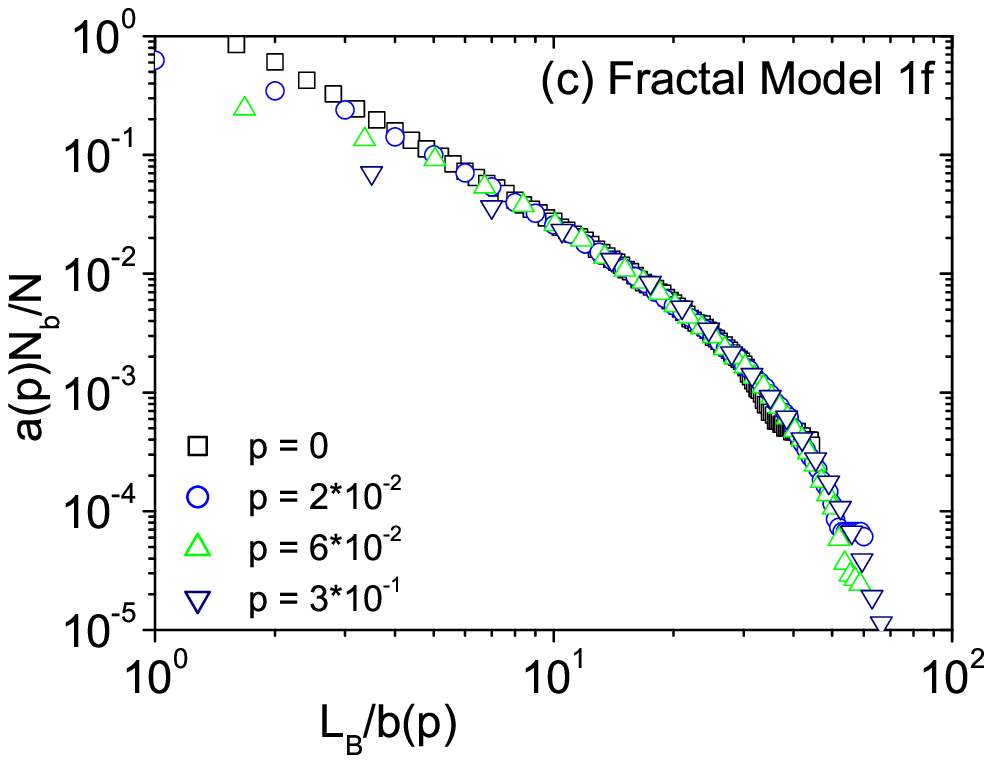}
\includegraphics[width=7.0 cm,height=5.0cm,angle=0]{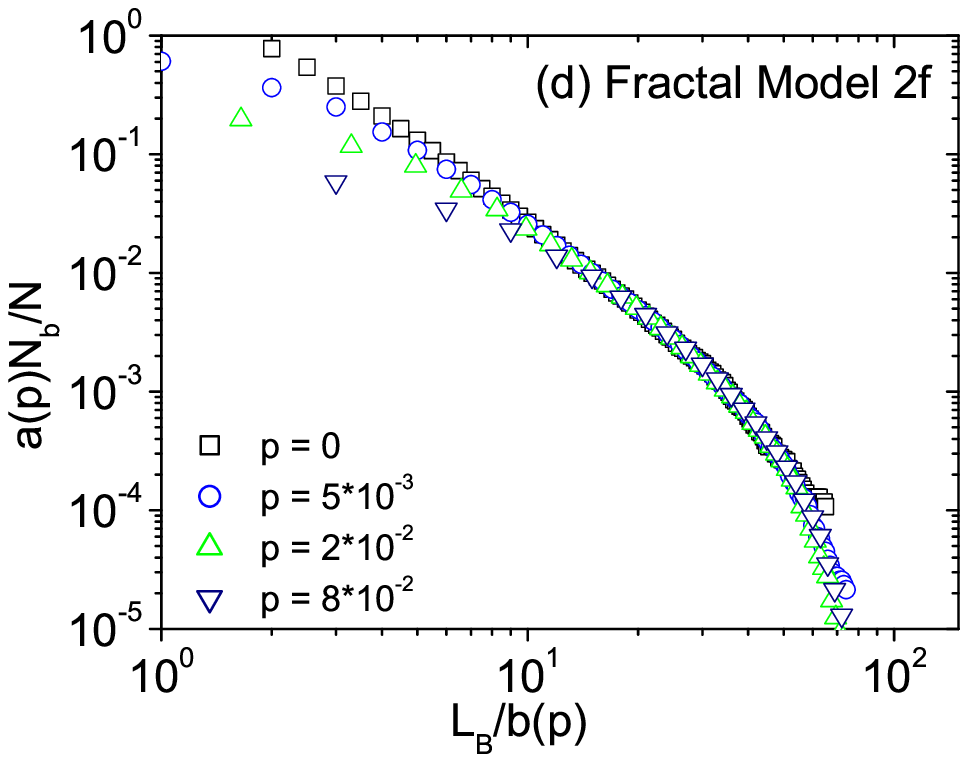}
\includegraphics[width=7.0 cm,height=5.0cm,angle=0]{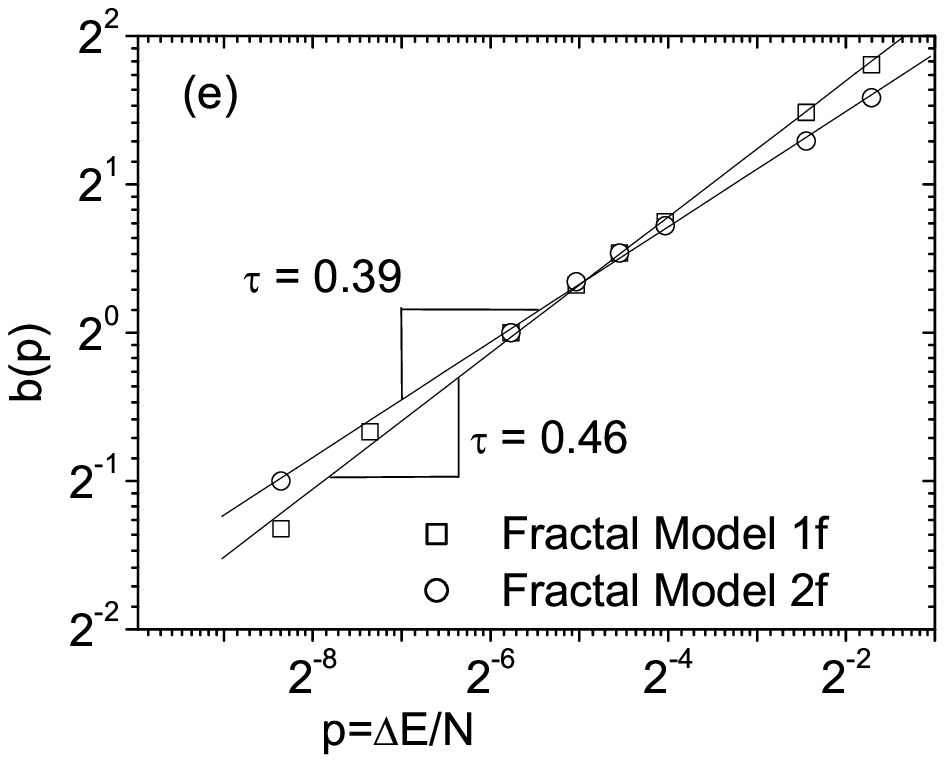}
\caption{(a,b) Box-covering analysis of fractal models 1f and 2f
with added random edges. Networks remain fractal for length scales
smaller than certain crossover length $\ell^{*}$. Above $\ell^{*}$
the networks are no longer fractals. The crossover length
$\ell^{*}$ becomes smaller as we add more edges. (c,d) Data
collapse of $N_{B}(\ell,p)$ for the two fractal models.
Appropriate rescaling $N_{B}(\ell_{B})\longrightarrow
a(p)N_{B}(\ell_{B}/b(p))$ allows to collapse all the values of
$N_{B}(l,p)$ onto a single curve. (e) The rescaling function
$b(p)\equiv \ell^{*}$ for fractal models 1f and 2f as a function
of $p$ shows a power law scaling of the crossover length $\ell^{*}
\sim p^{-\tau}$. Calculated exponents are $\tau_{1} = 0.46$ and
$\tau_{2} = 0.39$ respectively. Calculated values are slightly
smaller than the expected values due to finite size effects.}
\label{fig_scaling}
\end{figure}

We now study the behavior of fractal and non-fractal networks upon
adding random edges. We analyze the crossover from fractal to
non-fractal structure when random edges are added. To this end, we
study the minimal number of boxes $N_{B}$ of size $\ell_{B}$
needed to cover the network as a function of $\ell_{B}$ as we add
random edges to the network. Fig.~\ref{fig_scaling}(a) and
~\ref{fig_scaling}(b) show that the dimension $d_{B}$ of the
networks does not change. However, the network remains fractal
with a power law regime $N_{B}\sim \ell^{-d_{B}}$ only at length
scales $\ell$ below $\ell^{*}$, a characteristic length which
depends on $p$. For $\ell > \ell^{*}$, the network with added
random edges behaves as non-fractal with exponential decay
$N_{B}\sim\exp(-\ell/\ell^{*})$. The crossover length $\ell^{*}$
separating the fractal and non-fractal regions decreases as we add
more edges [see Figs.~\ref{fig_scaling}(a) and
\ref{fig_scaling}(b)]. We employ a scaling approach to deduce the
functional dependence of the crossover length on the density of
added shortcuts $p$. We propose for $N_{B}$ the scaling ansatz
\begin{equation}
N_B(\ell,p)\sim \ell^*(p)^{-d_B}F({\ell\over\ell^*(p)}),
\label{fig_scaling_ansatz}
\end{equation}
where
\begin{eqnarray}
   F(u) \sim \left
  \{\begin{array}{ll}
    u^{-d_B} & ~~~u \ll 1\\
    \exp(-u) & ~~~u \gg 1.
\end{array}\right.
\label{function}
\end{eqnarray}
With appropriate rescaling we can collapse all the values of
$N_{B}(\ell,p)$ onto a single curve [see Figs.~\ref{fig_scaling}(c)
and \ref{fig_scaling}(d)]. The crossover length $\ell^{*}(p)$
exhibits a clear power law dependence on the density of random
shortcuts [Fig.~\ref{fig_scaling}(e)],
\begin{equation}
\ell^{*}(p) \sim p^{-\tau}.
\end{equation}

We next argue that asymptotically for large $N$,
\begin{equation}
\tau = 1/d_{B}.
\end{equation}
When a fractal network with $N$ nodes and $E$ edges has additional
$\Delta \ll N$ random edges, the probability of a given node $i$
to have a random link is $P_{i}=2\Delta/N$. The mass of the
cluster within a size $\ell$ in a fractal network is $M_C\sim
\ell^{d_f}$. The probability of $M_{C}(\ell)$ possessing a random
edge is $P = (2\Delta/N)M_C$. Thus, at distances $\ell$ for which
$(2\Delta/N)M_C\ll 1$ we are in the {\it fractal\/} regime. On the
other hand, large distances $\ell$ for which $(2\Delta/N)M_C\gg 1$
correspond to the {\it non-fractal\/} regime. Thus, the crossover
length $\ell^{*}$ corresponds to $(2\Delta/N)M_C(\ell^{*})\sim 1$,
which implies $\ell^{*}\sim \Delta^{-1/d_{B}}$ or $\ell^{*}\sim
p^{-1/d_{B}}$, where $p\equiv \Delta/N$. Note that the values
measured for the two fractal networks, shown in Fig.
\ref{fig_scaling}(e), $\tau=0.46$~$(d_{B}=1.9)$ and
$\tau=0.39$~$(d_{B}=2.3)$, are slightly smaller then the expected
asymptotic values, which we attribute as likely to be due to
finite size effects.

\section{Discussion and Summary}

We have shown that node betweenness centrality and node degree are
significantly less correlated in fractal SF networks compared to
non-fractal SF networks due to the effect of repulsion between the
hubs. Betweenness centrality distribution in SF networks obeys a
power law $P(C) \sim C^{-\delta}$. We derived an analytic
expression for the betweenness centrality distribution exponent
$\delta = 2 - 1/d_{B}$ for SF fractal trees. Hence, fractal
networks with smaller dimension $d_{B}$ have more nodes with
higher betweenness centrality compared to networks with larger
$d_{B}$. The transition from fractal to non-fractal behavior was
studied by adding random edges to the fractal network. We observed
a crossover from fractal to non-fractal regimes at a crossover
length $\ell^{*}$. We found both analytically and numerically that
$\ell^{*}$ scales with density of random edges $p$ as $\ell^{*}
\sim p^{-\tau}$ with $\tau=1/d_B$.

\section{Acknowledgements}

We thank ONR, European NEST, project DYSONET and Israel Foundation
of Science for financial support. We are grateful to M. Riccaboni,
O. Penner, and S. Sreenivasan for helpful discussions.

\appendix

\section{A fractal growth model}

\begin{figure}[!h]
\includegraphics[width=1.0 cm,height=1.0cm,angle=0]{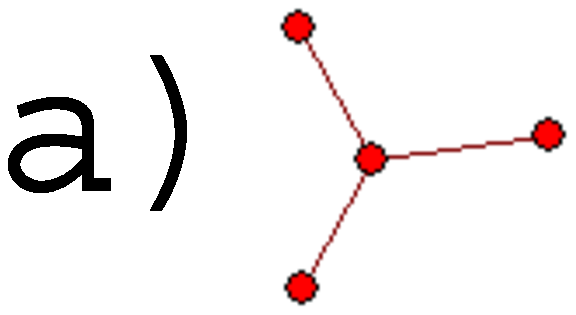}
\includegraphics[width=1.2 cm,height=1.2cm,angle=0]{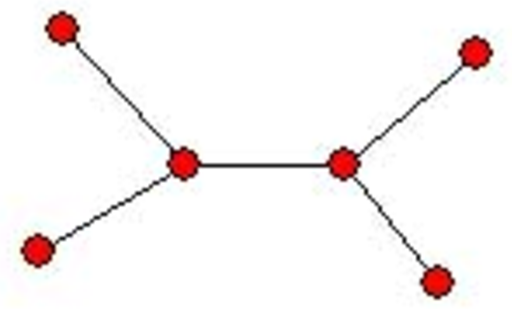}
\includegraphics[width=2.5 cm,height=2.5cm,angle=0]{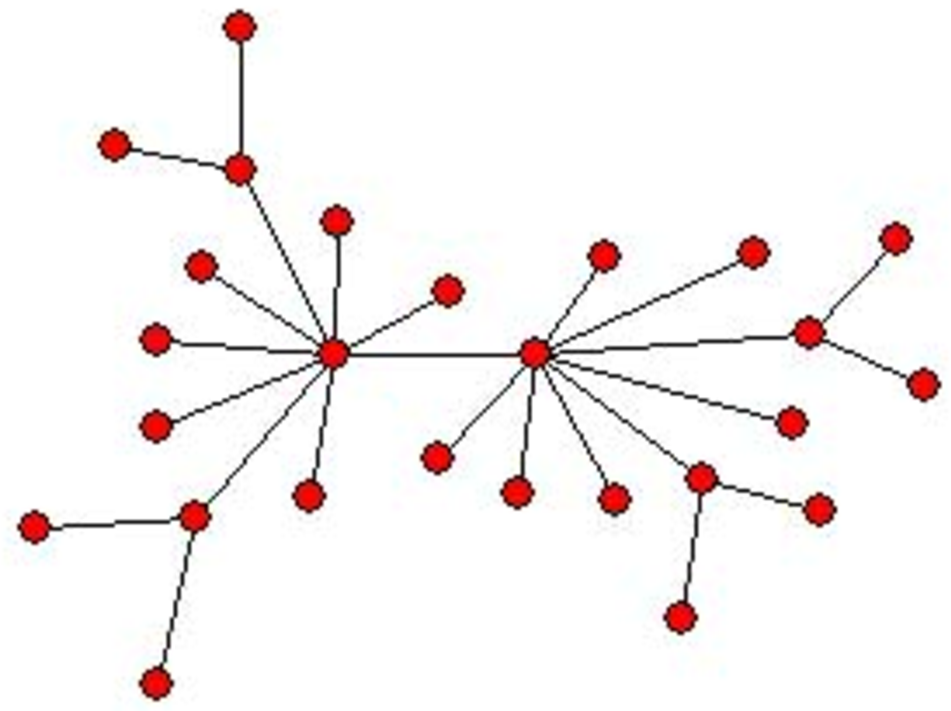}
\includegraphics[width=3.0 cm,height=3.0cm,angle=0]{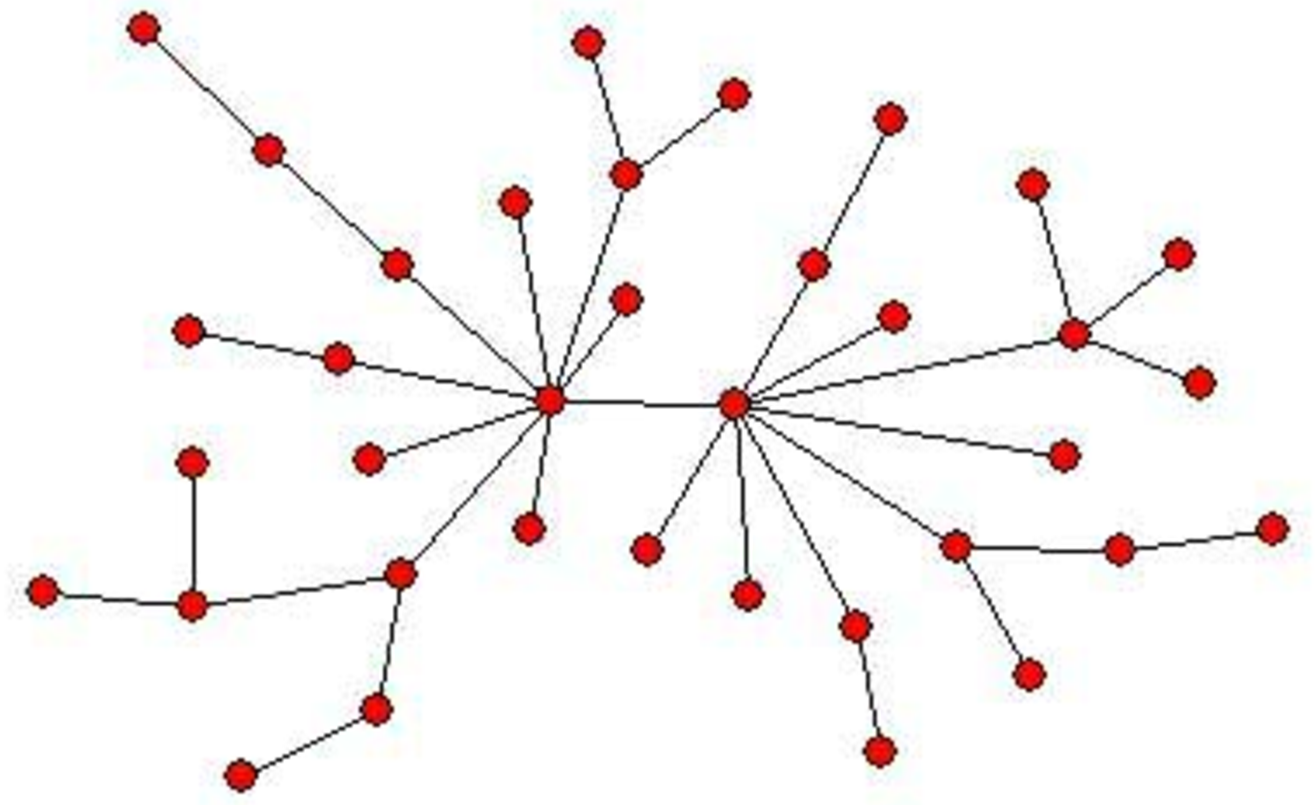}
\includegraphics[width=3.5 cm,height=3.5cm,angle=0]{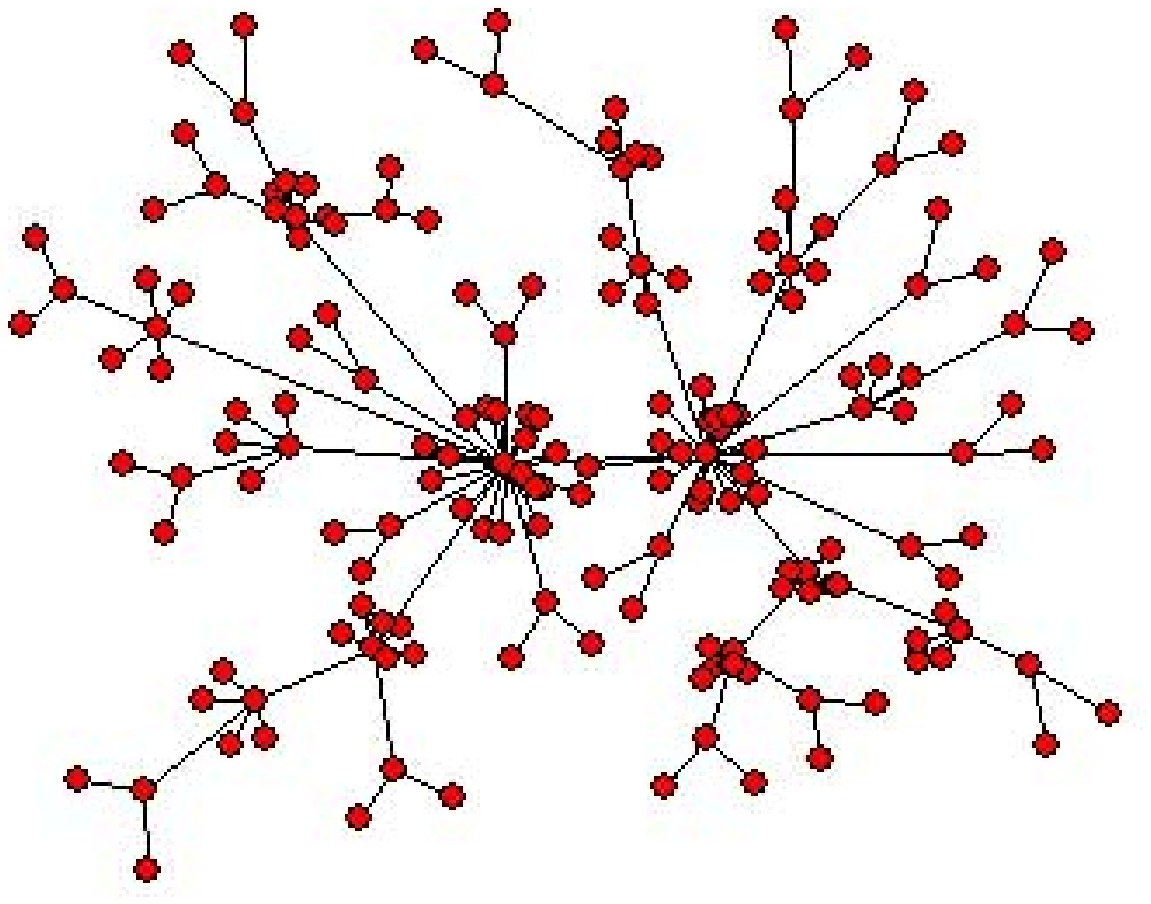}
\includegraphics[width=4.0 cm,height=4.0cm,angle=0]{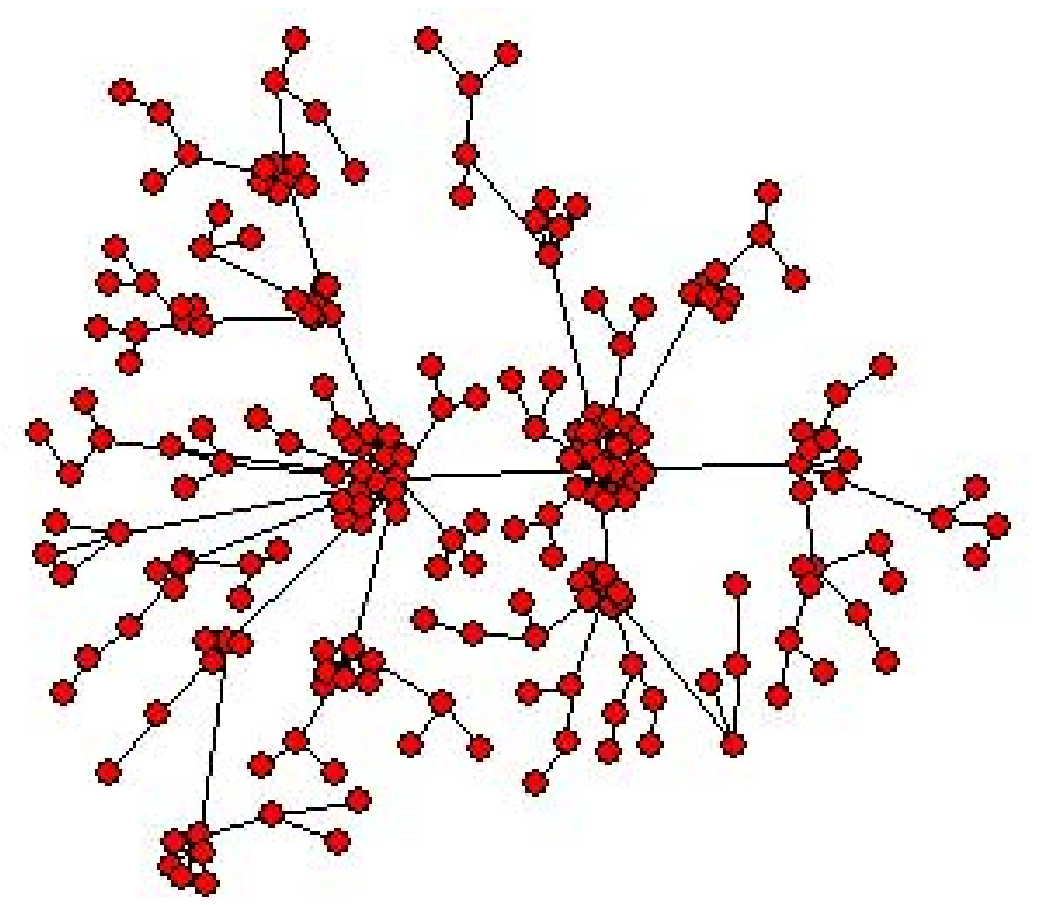}
\includegraphics[width=1.0 cm,height=1.0cm,angle=0]{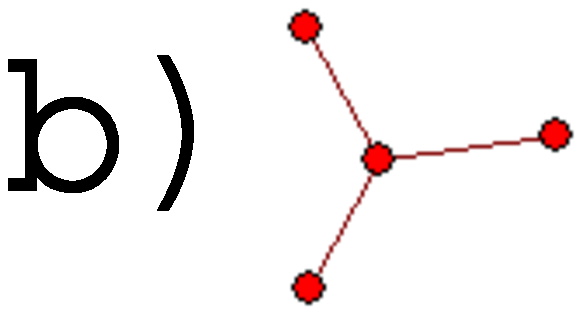}
\includegraphics[width=1.2 cm,height=1.2cm,angle=0]{fig6b.eps}
\includegraphics[width=2.5 cm,height=2.5cm,angle=0]{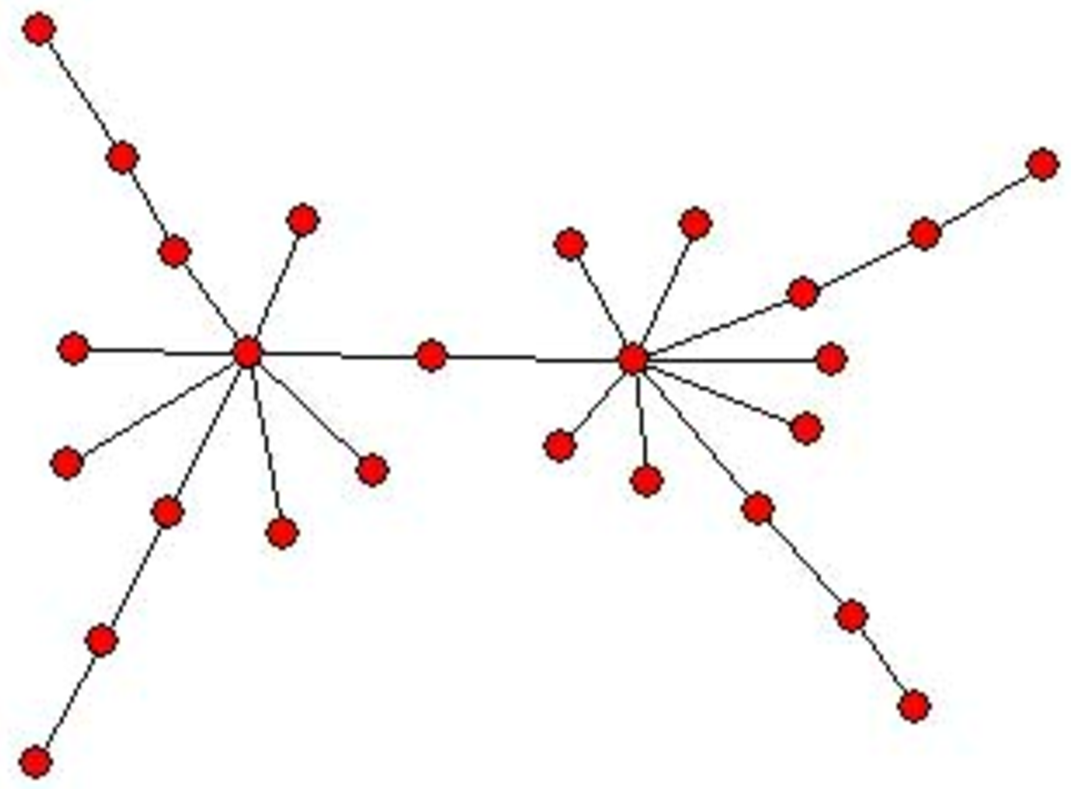}
\includegraphics[width=3.0 cm,height=3.0cm,angle=0]{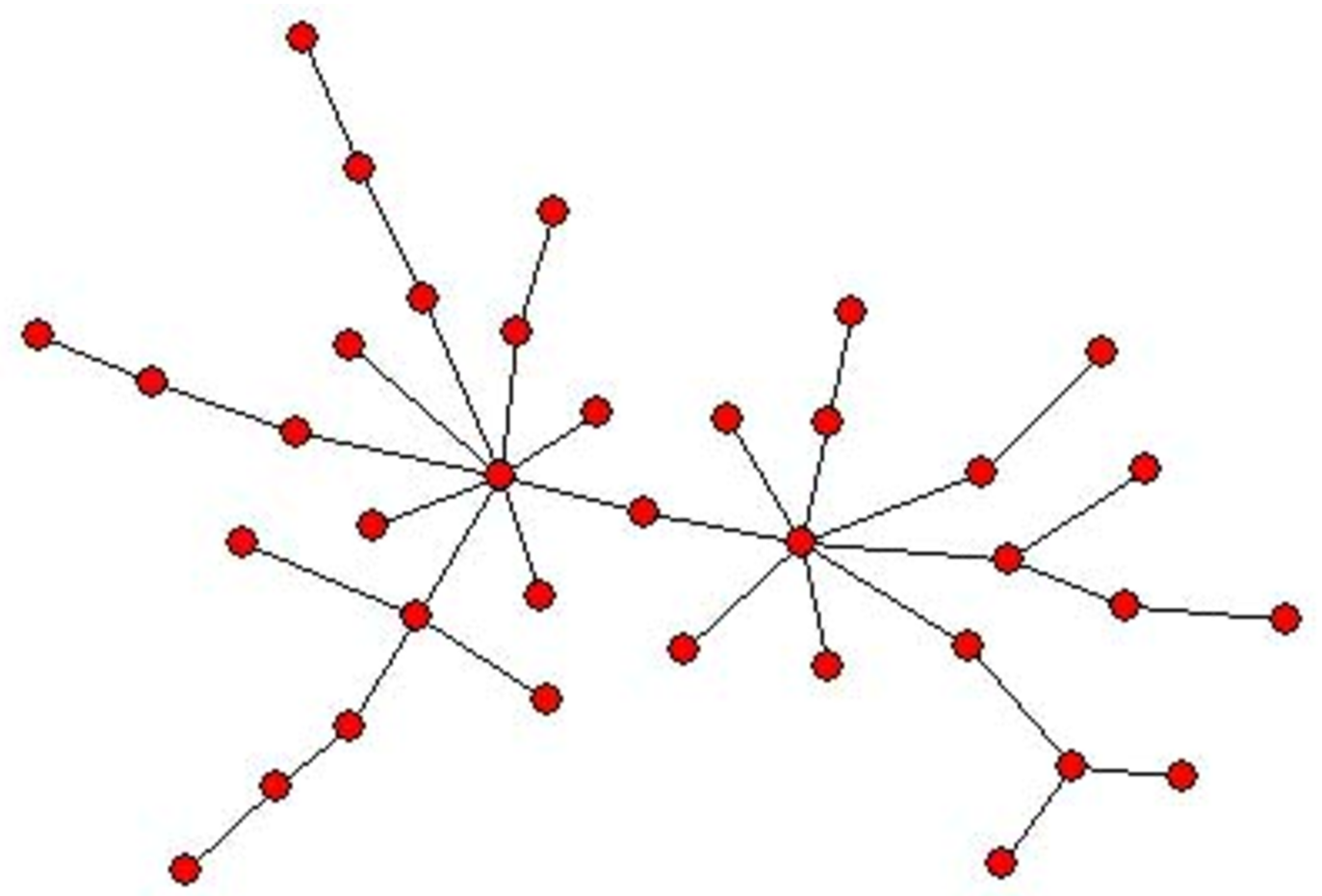}
\includegraphics[width=3.5 cm,height=3.5cm,angle=0]{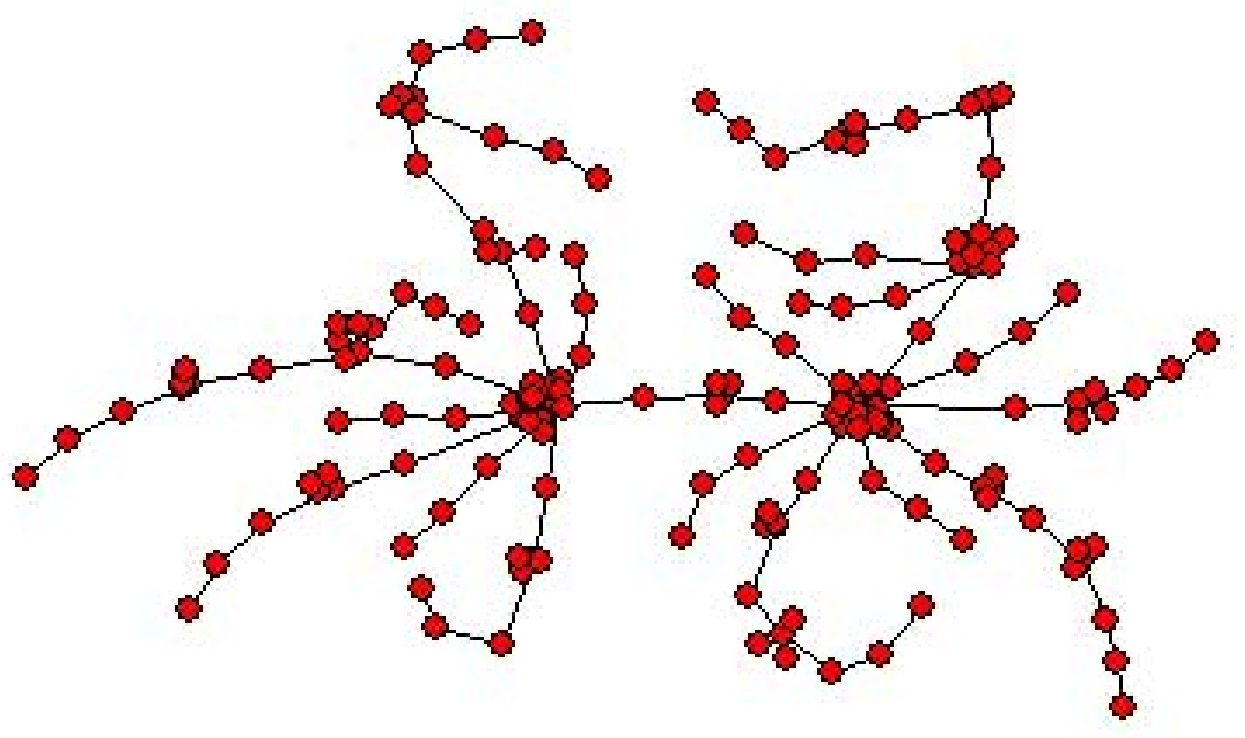}
\includegraphics[width=4.0 cm,height=4.0cm,angle=0]{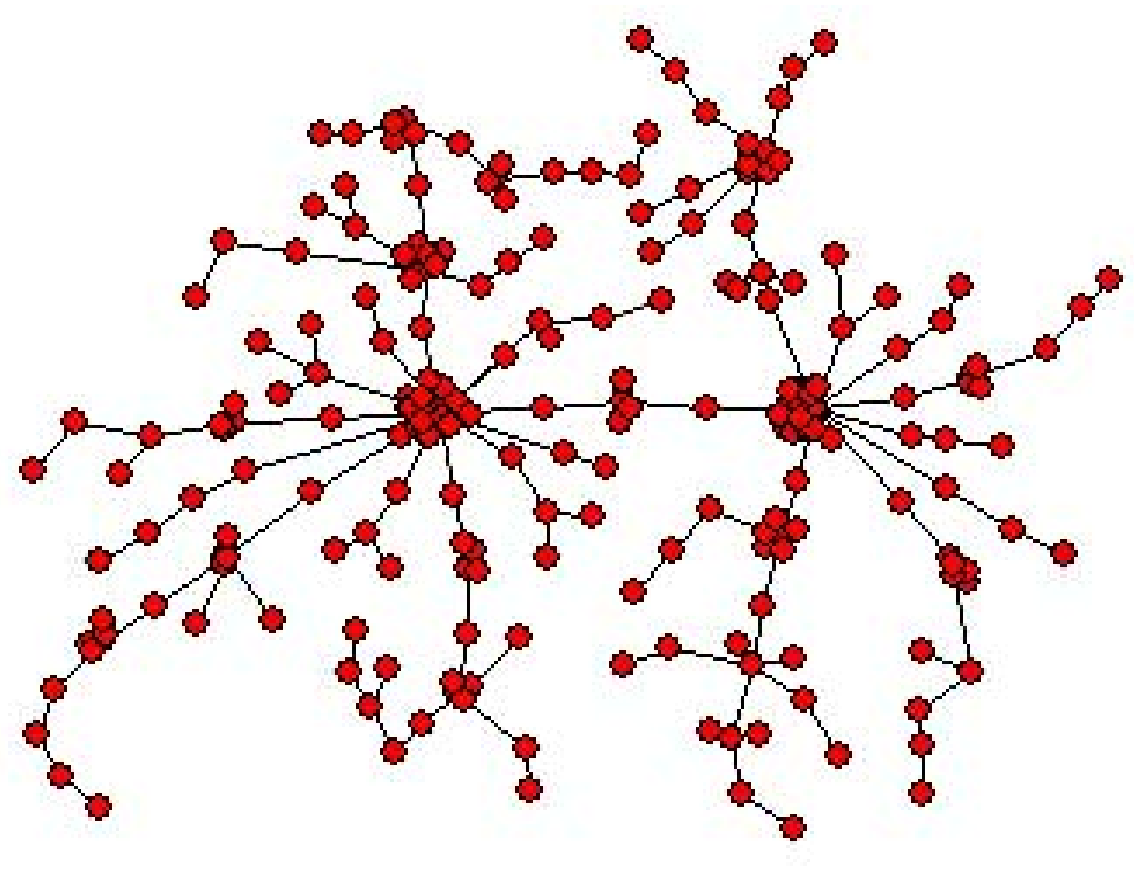}
\caption{Constructing (a) non-fractal and (b) fractal networks
with parameters $n=6$, $m=3$. The key difference between a fractal
and non-fractal model is ``repulsion between hubs''. In fractal
networks nodes of large degree prefer to connect to nodes of small
degree and not to nodes of large degree.}
\label{fig_growth_process}
\end{figure}

A growth model of fractal SF networks was first introduced by Song
et al. \cite{repulsion}. In the core of the growth model lies the
network renormalization technique \cite{self-sim, repulsion}: A
network is covered with $N_{B}$ boxes of size $\ell_{B}$.
Subsequently, each of the boxes is replaced by a node to construct
the renormalized network. The process is repeated until the network
is reduced to a single node. The fractal growth model represents the
inverse of this renormalization process. The growth process is
controlled by three parameters: $n$, $m$ and $e$ so that:
\begin{eqnarray}
N(t) &=& n*N(t-1) \label{growth1}
\\
k_{i}(t) &=& m*k_{i}(t-1), \label{growth2}
\end{eqnarray}
where $N(t)$ and $k_{i}(t)$ are, respectively, the number of nodes
of the network and degree of node $i$ at time $t$. The parameter $e$
is the probability of hub attraction $e\equiv E_{\rm hubs}/E$. In
the present study we limit our consideration to two distinct types
of models: fractal ($e=0$) and non-fractal ($e=1$). At each growth
step we run through all existing nodes. With probability $e$ we
increase the degree of a given node by attaching $(m-1)k_{i}(t-1)$
new nodes (this corresponds to hub attraction). With probability
$1-e$ we grow $(m-1)k_{i}(t-1) - 1$ nodes using remaining node to
repel hubs. Thus, the entire growth process can be summarized as
follows (see Fig.~\ref{fig_growth_process}):

\begin{itemize}

\item[{(1)}] Start with a single node

\item[{(2)}] Connect $(m-1)k_{i}(t-1)$ extra nodes to each node $i$ to
satisfy Eq.~(\ref{growth2}). With probability $1-e$ use one of the
new nodes to repel node $i$ from the central node.

\item[{(3)}] Attach the remaining number of nodes to the network
randomly to satisfy Eq.~(\ref{growth1}).

\item[{(4)}] Repeat steps (2) and (3) for the desired number of
generations $g$.

\end{itemize}

The networks constructed in this way are SF with
\begin{equation}
\lambda = 1 + {\log n\over\log m}. \label{SF_lambda}
\end{equation}
Fractal networks have a finite dimension
\begin{equation}
d_{B} = {\log n\over\log 2}. \label{dimensionality}
\end{equation}
For derivations of Eqs.~(\ref{SF_lambda}) and
(\ref{dimensionality}) see Ref.~\cite{repulsion}.

Here we refer to network models using a set of numbers (g,n,m,e).
For example, the set $(4,5,3,0)$ should read as a 4th generation
($g=4$) fractal ($e=0$) network with $n=5$ and $m=3$. According to
the above growth process for this example, $(4,5,3,0)$,
$N=n^g=625$, $E=N-1=624$, $\lambda = 1 + \log n/\log m = 2.46$,
and $d_{B} = \log n/\log 2 = 2.32$.

\section{Modified box counting method.}

The box counting method is used to calculate the minimum number of
boxes $N_{B}$ of size $\ell_{B}$ needed to cover the entire
network of $N$ nodes. The size of the box, $\ell_{B}$, imposes a
constraint on the number of nodes that can be covered: all nodes
covered by the same box must be connected and the shortest path
between any pair of nodes in the box should not exceed $\ell_{B}$.
The most crucial and time-consuming part of the method is to find
the minimum out of all possible combinations of boxes. In the
present study we use an approximate method that allows to estimate
the number of boxes rather fast.

\begin{itemize}

\item[{(1)}] Choose a random node (seed) on the network.

\item[{(2)}] Mark a cluster of radius $\ell_{B}$ centered on the chosen
node.

\item[{(3)}] Choose another seed on the unmarked part of the network.

\item[{(4)}] Repeat steps (2) and (3) until the entire network is
covered. The total number of seeds $N'$ is an estimate of the
required number of boxes $N_{B}$.

\end{itemize}

We stress that the estimated number of clusters $N'$ is always less
than $N_{B}$, the minimal number of boxes needed to cover the entire
network. Indeed, the shortest path between any two seeds is greater
then the size of the box $\ell_{B}$. Thus, a box cannot contain more
than one seed, and in order to cover the whole network we need at
least $N'$ boxes.

Even though $N'$ is always less or equal to $N_{B}$, the estimate
may be good or poor based on the order we choose for the nodes. In
order to improve the estimation we compute many times $N'$
(typically 100--1000) and choose the maximum of all $N'$.

\begin{figure}[!ht]
\includegraphics[width=8.0 cm,height=7.0cm,angle=0]{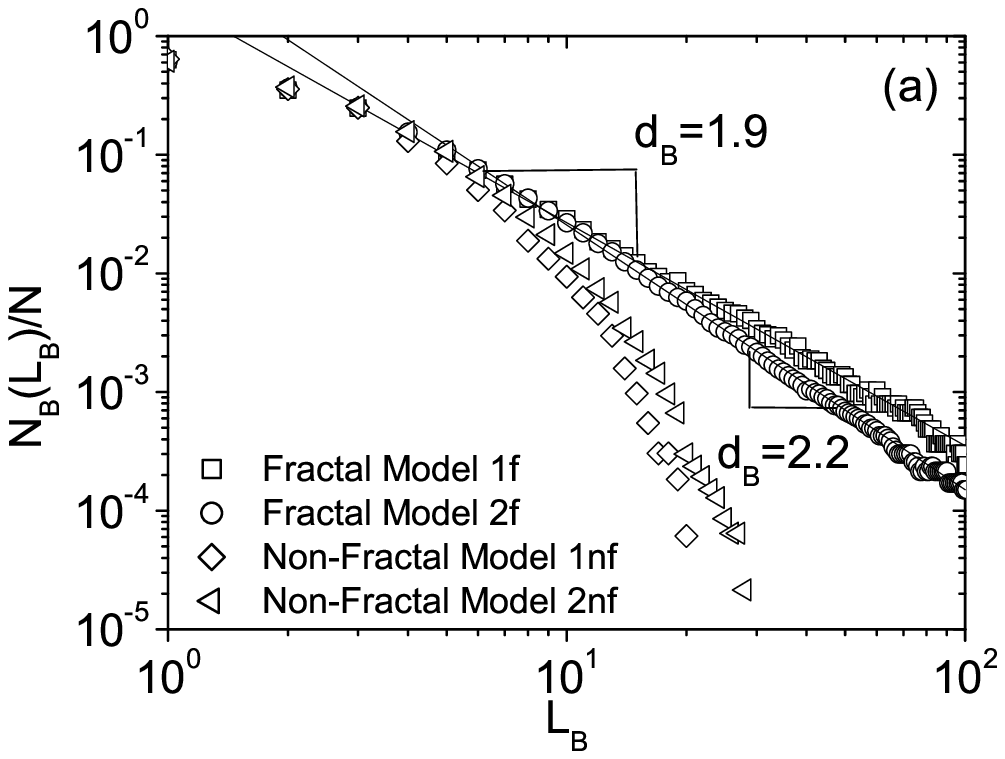}
\includegraphics[width=8.0 cm,height=7.0cm,angle=0]{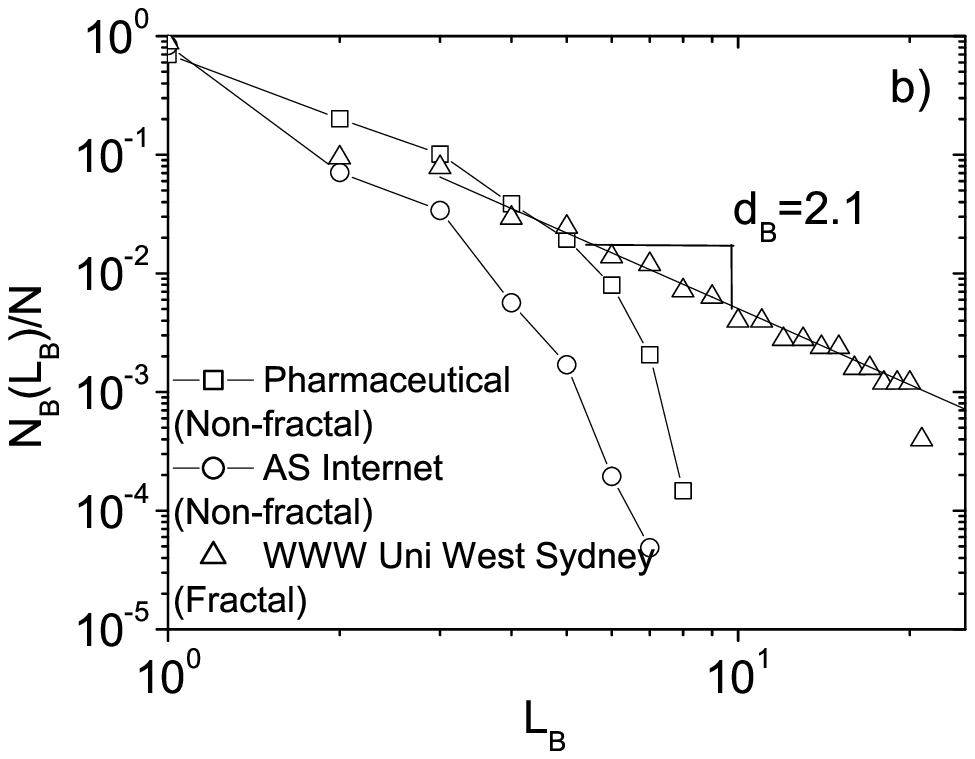}
\includegraphics[width=8.0 cm,height=7.0cm,angle=0]{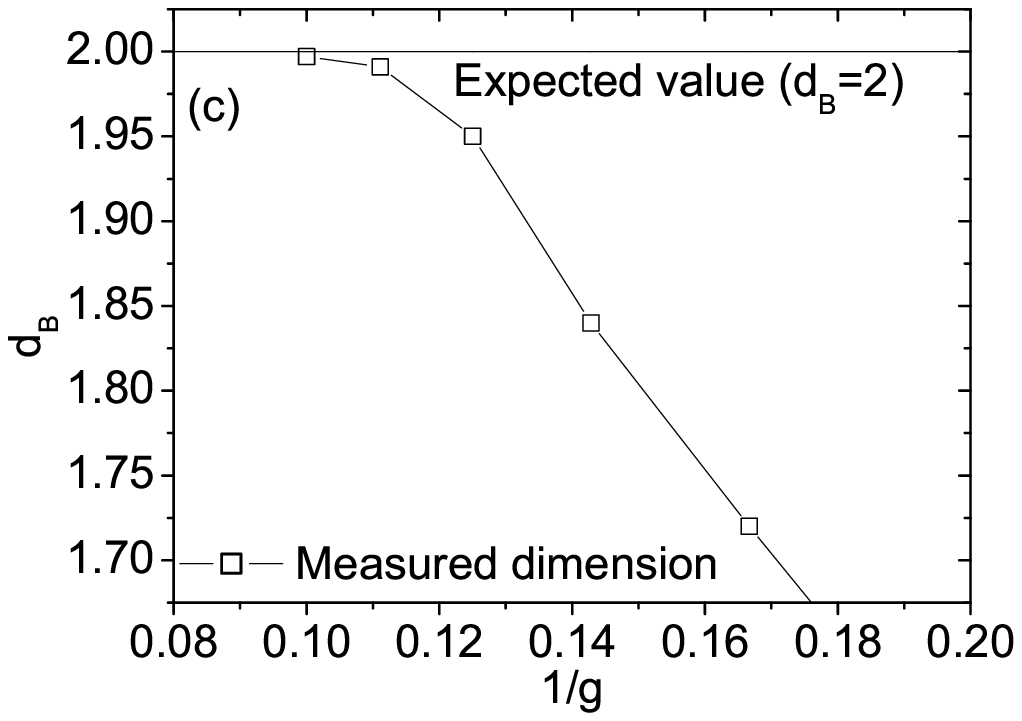}
\caption{Box-covering method applied to (a) models: 1f, 2f, 1nf
and 2nf and (b) real networks: WWW, Pharmaceutical, and AS
Internet. The log-log plots of the number of boxes $N_{B}$ needed
to cover the network as a function of their size $l_{B}$ show
clear ``power-law'' behavior for the fractal networks.  The
calculated dimensions are presented in Table I. (c) The calculated
dimension of fractal model 1f for different generations $g$ of the
same fractal model network. Calculated value of $d_{B}$ approaches
the expected value ($d_{B} = 2$) as the number of generations
increases.} \label{fig_nw_dims}
\end{figure}

Figures \ref{fig_nw_dims}(a) and \ref{fig_nw_dims}(b) demonstrate
the application of the modified box counting algorithm to several
fractal and non-fractal networks.  According to Eq.~(\ref{dim}),
dimensions of the fractal networks are obtained by calculating the
slope of the $N_{B}(\ell_{B})$ function in log-log format. The
calculated dimensions are slightly underestimated due to a finite
size effect of the analyzed networks.

Figure \ref{fig_nw_dims}(c) represents $d_{B}$ as a function of the
inverse number of generations $g$ of the model. As number of
generations $g$ increases calculated dimension $d_{B}$ approaches
the value given by Eq.~(\ref{dimensionality}).

A similar algorithm was introduced in
Ref.~\cite{crit_and_supercrit_skelet}. The authors of this
algorithm argue that it provides the same dimension of the network
no matter how the boxes are chosen. In our algorithm we intend to
estimate not only the dimension of the network but also the number
of boxes. Thus, we are seeking the maximum $N'$ out of many
realizations.

\newpage

\begin{table}
\begin{center}
\begin{tabular}
 {|c|c|c|c|c|c|c|} \hline
Network Name & N & E & $\lambda$ & $d_{B}$ & Category \\
\hline Model 1nf(7,4,2,1) \footnote{See Appendix A for
abbreviation.} & 16384 & 16383 & 3.0 & N/A  &
Non-Fractal \\
\hline Model 2nf(6,6,2,1) & 46656 & 46655 & 3.6 & N/A &
Non-Fractal \\
\hline Model 3nf(8,3,2,1) & 6561 & 6560 & 2.6 & N/A  &
Non-Fractal \\
\hline Model 1f(7,4,2,0) & 16384 & 16383 & 3.0 & 2. &
Fractal \\
\hline
Model 2f(6,6,2,0) & 46656 & 46655 & 3.6 & 2.6 & Fractal \\
\hline Model 3f(8,3,2,0) & 6561 & 6560 & 2.6 & 1.6 &
Fractal \\
\hline SF Model & 2668 & 3875 & 2.5 & N/A &
Non-Fractal \\
\hline
Uni West Sydney WWW  & 2526 & 4097 & 2.2 & 2.1 &Fractal \\
\hline
Pharmaceutical ~\cite{Pharm} & 6776 & 19801 & 2.4 & N/A & Non-Fractal\\
\hline
Yeast ~\cite{Barabasi} & 1458 & 1948 & 2.4 & 4.2 & Fractal \\
\hline
AS Internet ~\cite{dimes} & 20556 & 62920 & 2.1 & N/A & Non-Fractal \\
\hline
\end{tabular}

\caption{ Properties of the networks studied in the paper. Here
$N$ is the number of nodes, $E$ the number of edges, $\lambda$ the
degree distribution exponent ($P(k) \sim k^{-\lambda}$) and $d_B$
is the fractal dimension. The notation of model networks is
explained in Appendix A. We consider only the largest connected
cluster of the network if the original network is disconnected.}
\end{center}
\label{nw_table}
\end{table}

\begin{figure}[!ht]
\includegraphics[width=6.0 cm,height=5.0cm,angle=0]{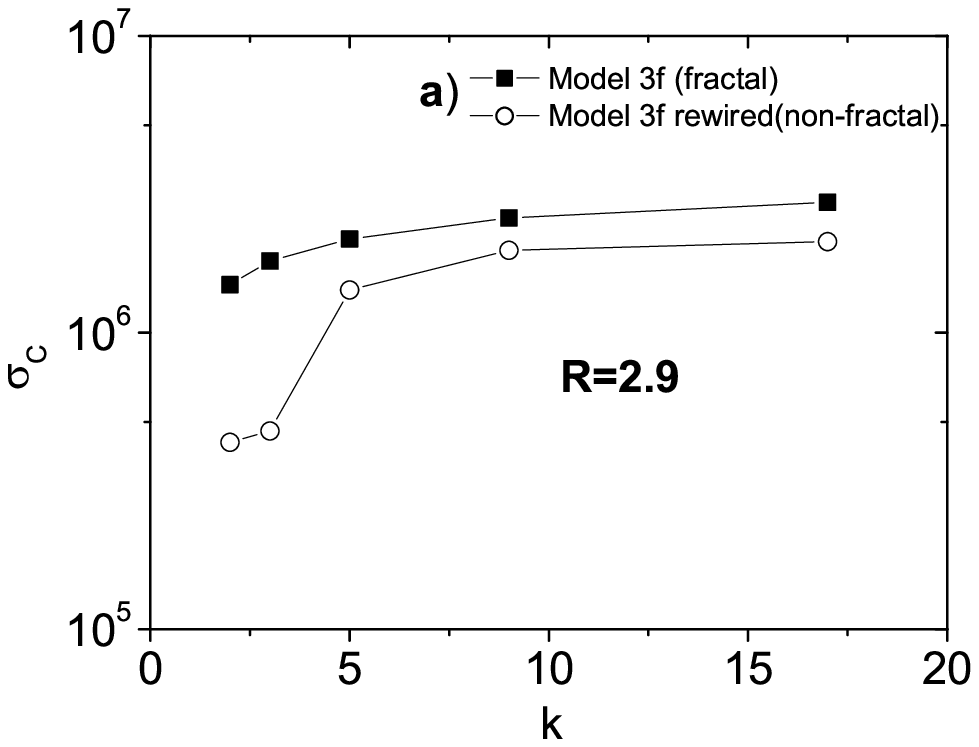}
\includegraphics[width=6.0 cm,height=5.0cm,angle=0]{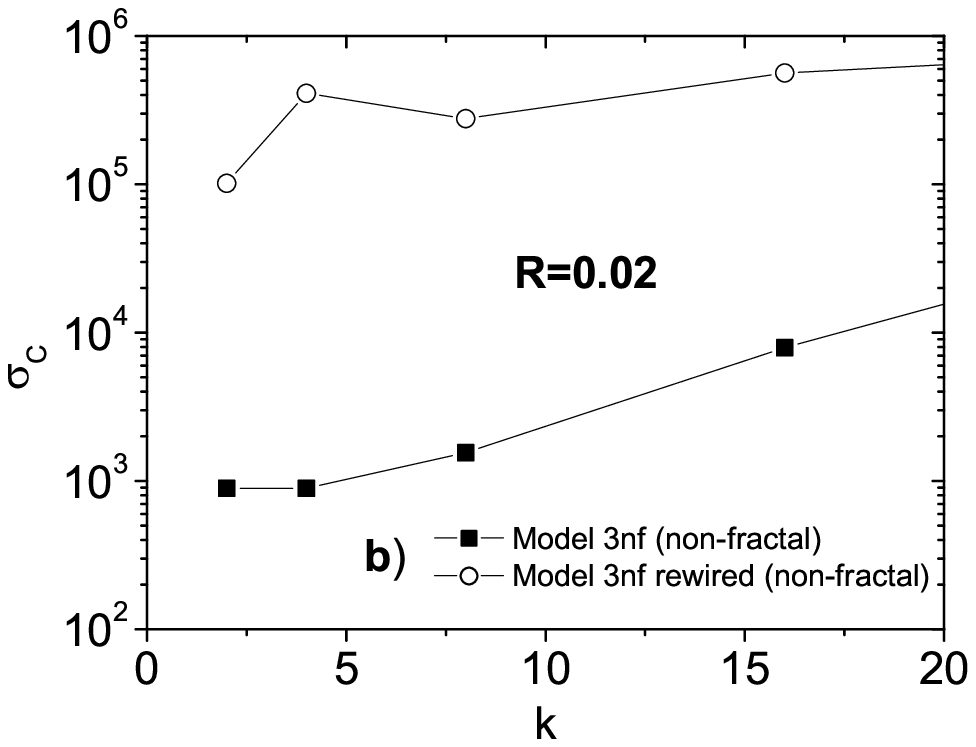}
\includegraphics[width=6.0 cm,height=5.0cm,angle=0]{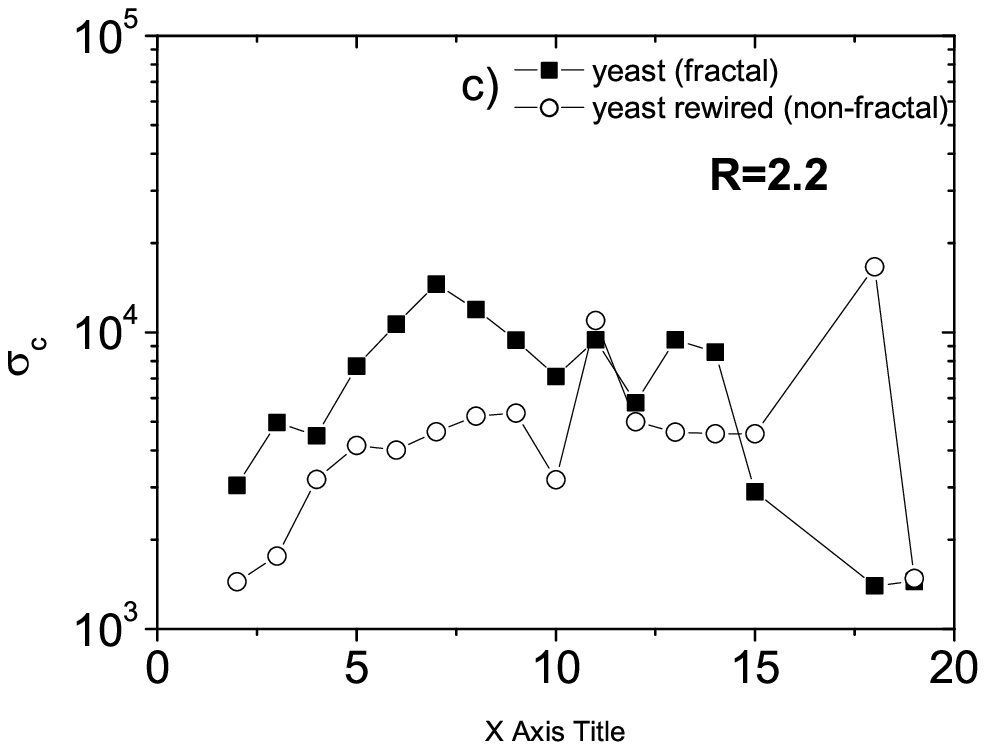}
\includegraphics[width=6.0 cm,height=5.0cm,angle=0]{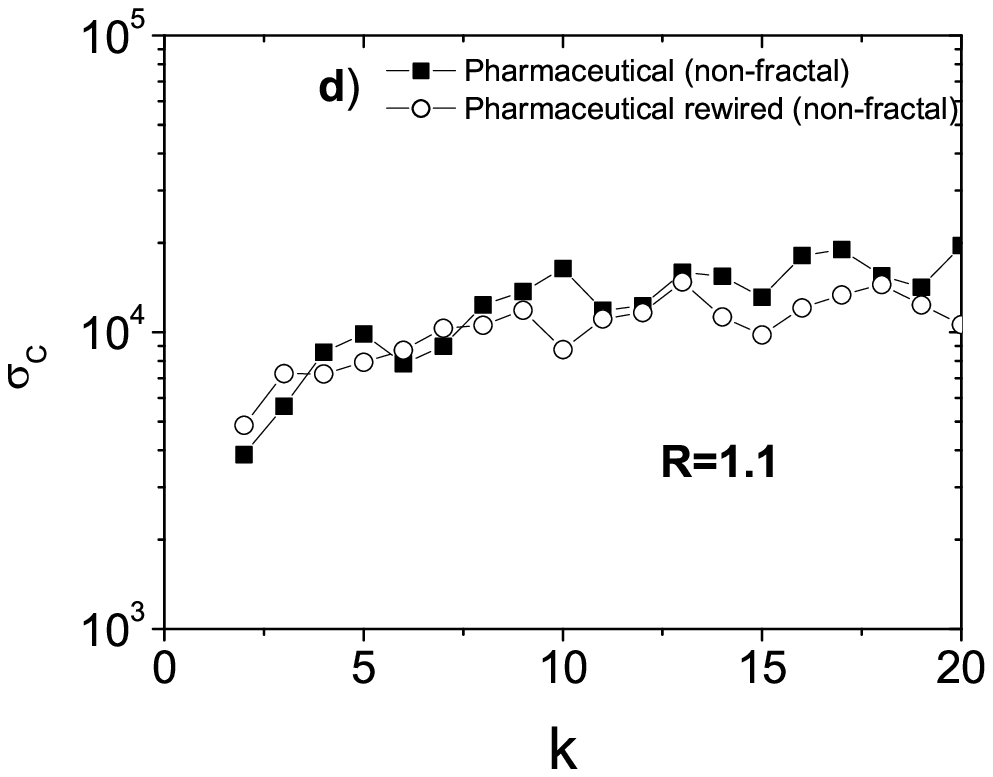}
\includegraphics[width=6.0 cm,height=5.0cm,angle=0]{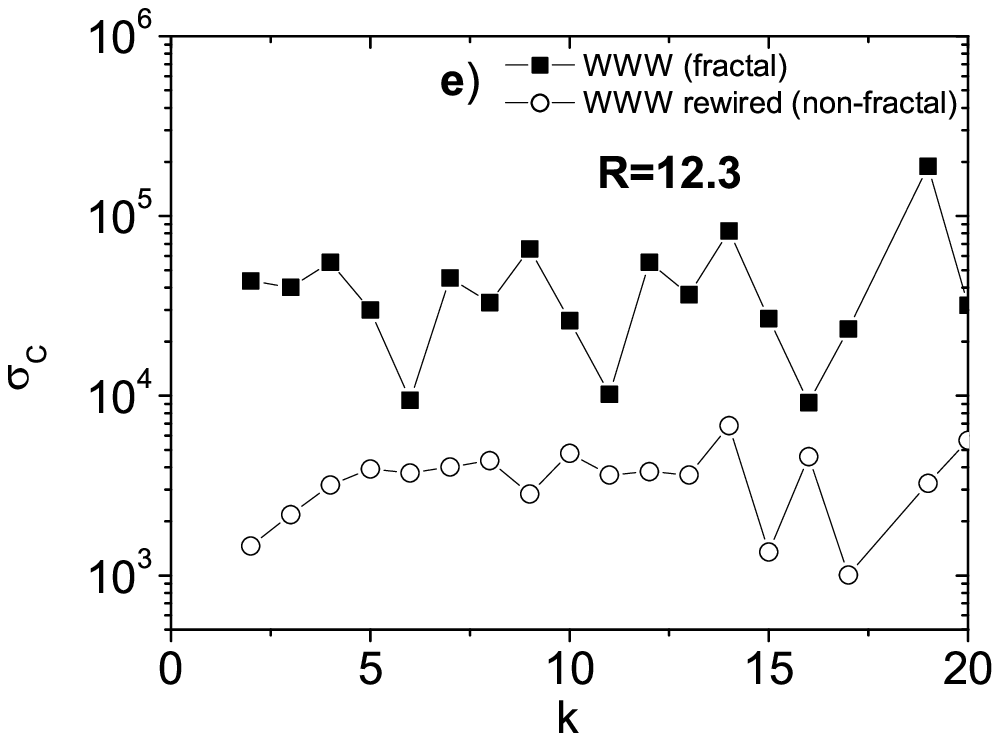}
\includegraphics[width=6.0 cm,height=5.0cm,angle=0]{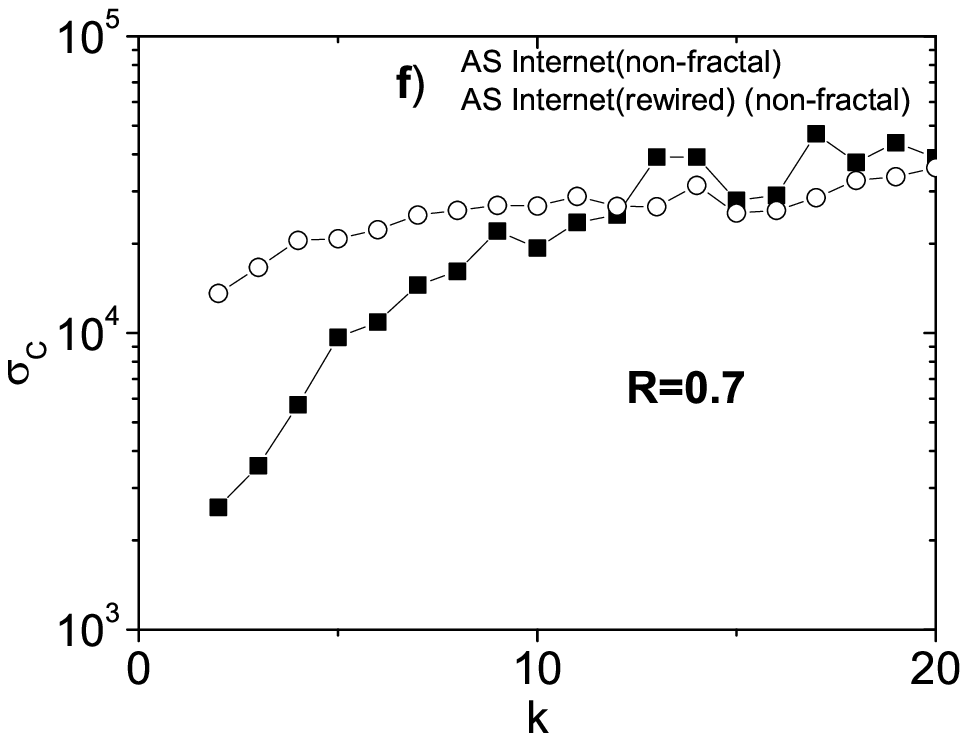}
\caption{Betweenness centrality variance $\sigma _{C}$ calculated
for both original and rewired networks as a function of node
degree $k$. Every point of the plot corresponds to the betweenness
centrality variance calculated for nodes of the \emph{same} degree
$k$ and normalized over the corresponding average betweenness
centrality value $<C>$ of the original network. Each of the plots
includes the value of the betweenness centrality- degree
correlation dispersion coefficient $R$, see Eq.(\ref{coeff}).
Note, that small degree nodes of fractal networks: fractal model
3f(a), yeast(c) and Uni Western Sydney WWW(e) have significantly
larger variance of betweenness centrality compared to their
randomly rewired counterparts which are non-fractals. On the other
hand, small degree nodes of the non-fractal networks: non-fractal
model 3nf(b), pharmaceutical(d) and AS internet(f) have
betweenness centrality variance comparable or even smaller than
that of their randomly rewired counterparts. As a result $R > 1$
for fractal networks and $R < 1$ or $R\approx1$ for nonfractal
networks. Thus, betweenness centrality - degree correlation is
weaker in fractal networks than in non-fractals.} \label{variance
profiles}
\end{figure}

\begin{thebibliography}{99}

\bibitem{Barabasi_rmp_review} R. Albert and A.-L. Barab\'{a}si
Rev. Mod. Phys. \textbf{74}, 47 (2002).

\bibitem{evol_of_nw} S. N. Dorogovtsev and J. F. F. Mendes,
\emph{Evolution of Networks: From Biological Nets to the Internet
and WWW\/} (Oxford University Press, Oxford, 2003).

\bibitem{evol_and_struct} R. Pastor-Satorras and A. Vespignani,
\emph{Evolution and Structure of the Internet: A Statistical Physics
Approach} (Cambridge University Press, Cambridge, 2004).

\bibitem{compl_nw} R. Cohen and S. Havlin (Cambridge University Press,
Cambridge,in press 2007).

\bibitem{ER} P. Erd\"{o}s and A. R\'{e}nyi,
Publ. Math. Inst. Hung. Acad. Sci. \textbf{6}, 290 (1959).

\bibitem{ER_random graphs} P. Erd\"{o}s and A. R\'{e}nyi,
Publ. Math. Inst. Hung. Acad. Sci. \textbf{5}, 17 (1960).

\bibitem{bollobas} B. Bollobas, \emph{Random Graphs\/} (Cambridge
University Press, 2001).

\bibitem{sw_problem} S. Milgram, Psychol. Today \textbf{2}, 60 (1967).

\bibitem{collect_dynam} D. J. Watts and S. H. Strogatz,
 Nature \textbf{393}, 440 (1998).

\bibitem{diam_of_www} R. Albert, H. Jeong, and A.-L. Barabasi,
Nature \textbf{401}, 130 (1999).

\bibitem{scaling_in_nw} A. L. Barab\'{a}si and R. Albert,
Science \textbf{286}, 509 (1999).

\bibitem{faloutsos_etall} M. Faloutsos, P. Faloutsos, and C. Faloutsos,
Comput. Comm. Rev. \textbf{29}, 251 (1999).

\bibitem{self-sim} C. Song, S. Havlin, and H. Makse,
Nature \textbf{433}, 392 (2005).

\bibitem{repulsion} C. Song, S. Havlin, and H. Makse,
Nature Physics \textbf{2}, 275 (2006).

\bibitem{skeleton_and_fractal} K. I. Goh, G. Salvi, B. Kahng, and D.
Kim, Phys. Rev. Lett. \textbf{96}, 018701 (2006).

\bibitem{crit_and_supercrit_skelet} J. S. Kim, K. I. Goh, G. Salvi,
E. Oh, B. Kahng, and D. Kim, cond-mat/0605324.

\bibitem{dissasort} S.-H. Yook, F. Radicci and H. Meyer-Ortmanns,
Phys. Rev. E. \textbf{72}, 045105(R) (2005).

\bibitem{box_count} J. Feder,
\emph{Fractals} (Plenum, New York, 1988).

\bibitem{assortativity} M.E.J. Newman, Phys. Rev. Lett.
\textbf{89}, 208701 (2002).

\bibitem{centr_definition} L. C. Freeman, Social Networks
\textbf{1}, 215 (1979).

\bibitem{centr_book} S. Wasserman and K. Faust, \emph{Social Network
Analysis\/} (Cambridge University Press, Cambridge, 1994)

\bibitem{centr_handbook} J. Scott, \emph{Social Network Analysis: A
Handbook\/} (Sage Publications, London, 2000)

\bibitem{newman} M. E. J. Newman,
Phys. Rev. E. \textbf{64}, 016132 (2001).

\bibitem{superhighways} Z. Wu, L. A. Braunstein, S. Havlin and H. E.
Stanley, Phys. Rev. Lett. \textbf{96}, 148702 (2006).

\bibitem{Pharm} L. Orsenigo, F. Pammolli, and M. Riccaboni,
Research Policy, \textbf{30(3)}, 485 (2001).

\bibitem{dimes}
{\tt http://netdimes.org} (The DIMES project).

\bibitem{Shai}
S. Carmi, S. Havlin, S. Kirkpatrick, Y. Shavitt and E. Shir,
cs.NI/0607080 (2006).

\bibitem{Barabasi}
{\tt http://www.nd.edu/$\sim$alb/} (Home page of A. L.
Barab\'{a}si).

\bibitem{yeast}
H. Jeong, S. Mason, A.-L. Barab\'{a}si), Z.N.Oltvai, Nature \
\textbf{411}, 41 (2001).

\bibitem{uws}
{\tt http://cybermetrics.wlv.ac.uk/database/} (The Academic Web
Link Database Project).

\bibitem{Attack_vulnerability} P. Holme, B. J. Kim, C. N. Yoon, and S.
K. Han, Phys. Rev. E \textbf{65}, 056109 (2002).

\bibitem{centr_distr_1} D. H. Kim, J.D. Noh, and H. Jeong,
Phys. Rev. E \textbf{70}, 046126 (2004).

\bibitem{centr_distr_2} K. I. Goh, J. D. Noh, B. Kahng, and D. Kim,
Phys. Rev. E \textbf{72}, 017102 (2005).

\bibitem{Braunstein_review} L. A. Braunstein, Z. Wu, T. Kalisky,
  Y. Chen, S. Sreenivasan, R. Cohen, E. L\'opez, S. V. Buldyrev,
  S. Havlin, and H. E. Stanley, ``Optimal Path and Minimal Spanning
  Trees in Random Weighted Networks'', Journal of Bifurcation and
Chaos
  {\bf xx}, xxx--xxx (2006).  cond-mat/0606338.

\bibitem{Bunde_Havlin} A. Bunde and S. Havlin, eds., \emph{Fractals in
Science} (Springer, Berlin, 1996).

\end{thebibliography}
\end{document}